# Millimeter-Wave Gaussian-to-Flat-Top Beam Shaping Using a Single-Curvature Lens: A Novel Approach

**Ehsan Imanbeygi**
M.Sc. Student, Department of Electrical Engineering, Sharif University of Technology, Tehran, Iran

**Mohammad Hossein Koohi Ghamsari**
M.Sc. Student, Department of Electrical Engineering, Sharif University of Technology, Tehran, Iran

**Mehdi Ahmadi-Boroujeni**
Assistant Professor, Department of Electrical Engineering, Sharif University of Technology, Tehran, Iran

**Abstract**
This paper presents a novel method for converting the divergent Gaussian intensity distribution of a millimeter-wave source into a uniform (flat-top) distribution using a single curved surface. The proposed method combines the surface profile of a collimator with that of a beam shaper. To validate the method, the profile of a single-sided (plano-convex) lens is derived and placed 52 mm in front of a conical horn antenna operating at 100 GHz. The lens transforms the Gaussian radiation of the horn, with an incident beam radius of 11.9 mm, into a uniform beam with a radius of 400 mm at a distance of 3 m. The lens surface profile is optimized using a full-wave-simulation-based procedure, and the resulting output beam is compared with that of a previously designed double-sided (concave-convex) lens. Simulation results show that, within the desired field of view, the optimized single-sided lens produces a flat-top intensity distribution with satisfactory ripple and roll-off. Finally, the sensitivity of the optimized lens to frequency variation is investigated. The proposed lens is suitable for generating uniform illumination in active stand-off millimeter-wave imaging systems.

Corresponding author: Mehdi Ahmadi-Boroujeni
Corresponding author's e-mail: ahmadi@sharif.edu

## 1- Introduction

The millimeter-wave region spans wavelengths from 1 to 10 mm and overlaps with the terahertz band at wavelengths of approximately 1 to 3 mm [1]. Over the past decade, considerable attention has been directed toward millimeter-wave technology for a wide range of applications, particularly imaging, because electromagnetic waves in this spectral region interact favorably with many materials. For example, many nonmetallic materials, including polymers and semiconductors, are transparent to electromagnetic radiation in this range. Moreover, because of the low photon energy, millimeter-wave radiation is non-ionizing. Owing to these characteristics, millimeter-wave imaging has found applications in medical imaging, security screening, and reflection and transmission spectroscopy [2]-[5].

In an active stand-off imaging system, the transmitter and receiver are located on the same side of the target, and an external source is used to illuminate the target. The image is then reconstructed by analyzing the waves scattered back from the target at a stand-off distance [6], [7]. The design of such systems involves several challenges, including transmitter beam shaping and improvement of imaging quality [8]. Beam shaping refers to modifying the spatial distribution of beam intensity into a desired profile. The external source used in such a system radiates a Gaussian intensity distribution; part of the radiated power spills beyond the target plane, while the remainder illuminates the field of view. Spillover efficiency is defined here as the ratio of the spilled power to the total radiated power, whereas Gaussian-intensity nonuniformity is defined as the ratio of the maximum Gaussian intensity to the Gaussian intensity at the edge of the field of view. Figure 1 plots the spillover efficiency as a function of Gaussian-intensity nonuniformity in decibels. As shown, reducing the spilled power increases the nonuniformity of the Gaussian distribution, causing the center of the target plane to be illuminated more strongly than its edges and thereby degrading image quality. To address this problem, a beam shaper is used to transform the Gaussian intensity into a uniform distribution so that the target plane is illuminated uniformly [9]. Beyond imaging, beam shapers are also used in a variety of optical applications, including material processing [10], micromachining [11], optical-signal processing [12], holography [13], and camera optics [14].

In [15], a Gerchberg-Saxton algorithm based on an iterative optimization procedure was used to derive the surface profile of a plano-convex refractive lens. A polypropylene refractive lens was then employed in an active stand-off terahertz imaging system to transform the Gaussian intensity radiated by a diagonal horn antenna operating at 625 GHz into a uniform intensity distribution at a distance of 3 m from the lens. In [16], Fermat's principle was first used to derive the surface profile of a collimating plano-convex lens; a geometrical-optics transformation method was then applied to derive the surface profile of a concave-plano lens for beam shaping. Combining the two derived lenses produced a concave-convex, or double-sided, lens. A silicon double-sided lens with a refractive index of 1.45 was used to transform the Gaussian intensity of a conical horn antenna operating at 100 GHz into a uniform intensity distribution 3 m from the lens. In the present work, the surface profiles obtained from the geometrical-optics transformation method and Fermat's principle are combined to design a plano-convex lens surface. Because the resulting lens has curvature on only one side, it is referred to here as a single-sided lens. Relatively little work has investigated single-sided millimeter-wave beam-shaping lenses using this approach. The proposed method results in a simpler structure and, despite its simplicity, is novel and produces the desired flat-top intensity distribution. The designed lens is used to transform the conical-horn beam into a uniform intensity distribution at a distance of 3 m, and the previously reported double-sided lens is used as a benchmark for validating the single-sided design.

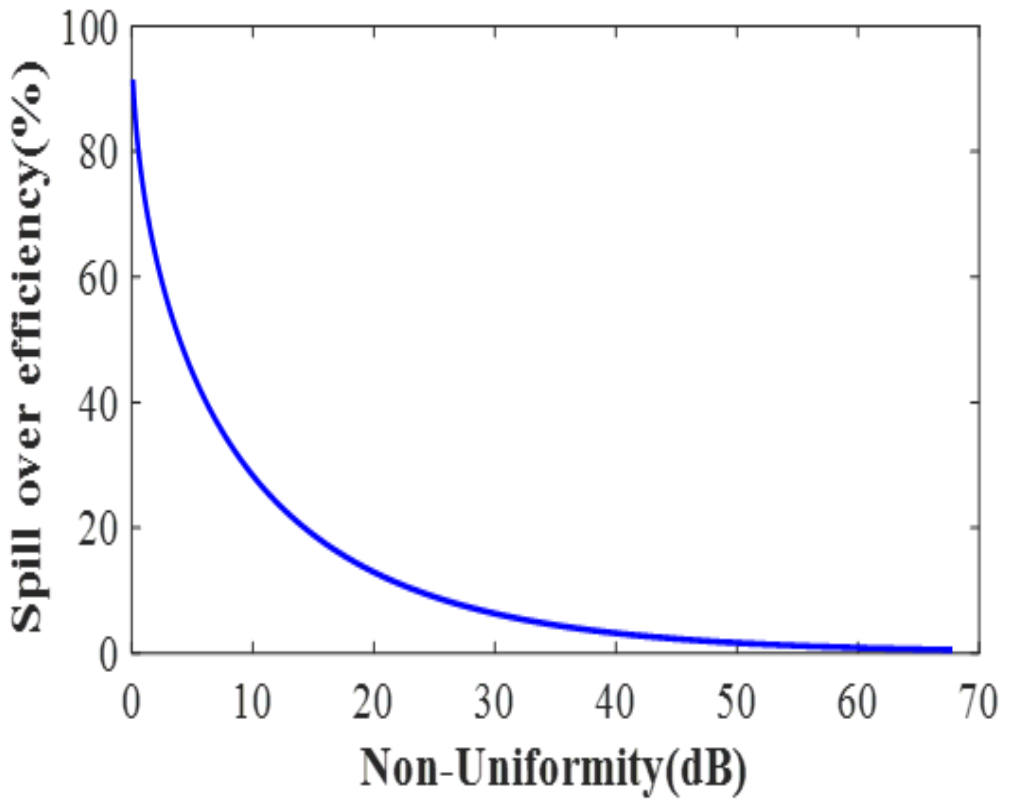

**Fig. 1. Spillover efficiency versus Gaussian-beam intensity nonuniformity.**

Section 2 first introduces the physical characteristics of the conical horn antenna used as the millimeter-wave source. In Section 3, a new method is employed to convert the double-sided concept into a single-sided lens, and the single-sided surface profile is designed based on the underlying beam-shaping physics. Section 4 presents full-wave simulations of the lens, compares its output intensity with that of the double-sided lens, and introduces performance metrics for assessing the quality of the resulting flat-top intensity distribution. The sensitivity of the single-sided-lens output to frequency variation is then examined and compared with that of the double-sided lens. Finally, Section 5 summarizes the main results for the single-sided lens.

## 2- Conical Horn Antenna

The two-dimensional schematic of the horn antenna used as the millimeter-wave source is shown in Fig. 2. The cylindrical waveguide is coupled to free space through a truncated cone. The waveguide operates at 100 GHz and has a cutoff frequency of 92.5 GHz. Only the dominant $TE_{11}$ mode propagates in the waveguide. The antenna gain, half-power beamwidth (HPBW) in the $\varphi = 0^{\circ}$ plane, and the distance from the antenna phase center to its outer aperture $\left(d_{pc}\right)$, together with the principal geometrical parameters - the outer-aperture diameter $\left(D_h\right)$, inner-aperture diameter $\left(D_{wg}\right)$, horn length $\left(L_h\right)$, and waveguide length $\left(L_{wg}\right)$ - are listed in Table 1.

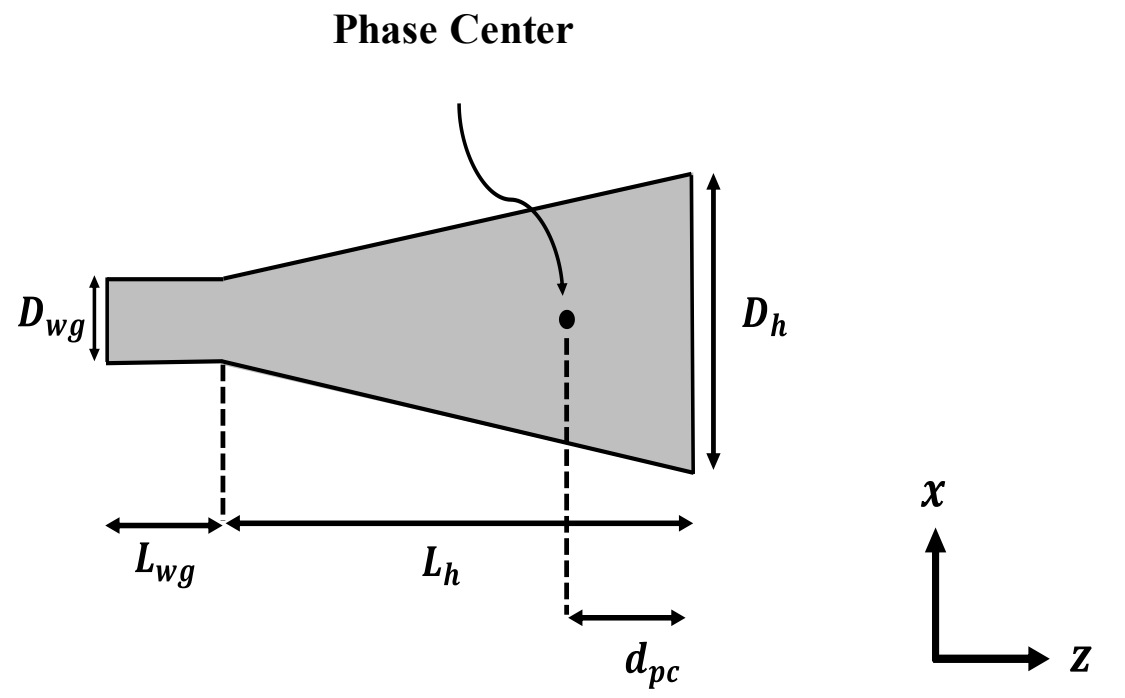

**Fig. 2. Two-dimensional schematic of the conical horn antenna.**

**Table 1. Geometrical dimensions and principal parameters of the conical horn antenna.**

| $L_{wg}$ | $L_h$ | $D_{wg}$ | $D_h$ | $d_{pc}$ | HPBW | gain |
|---|---|---|---|---|---|---|
| 3mm | 27mm | 1.9mm | 12mm | 9mm | 15.4° | 19.7dBi |

## 3- Design of the Single-Sided (Plano-Convex) Lens Surface Profile

Figure 3 shows a two-dimensional schematic of the single-sided dielectric lens and the source. As illustrated, the divergent rays radiated from the antenna phase center with a Gaussian intensity distribution impinge on the lens. Each ray travels along a different optical path through the lens and is deflected by a prescribed amount, thereby transforming the spatial beam-intensity distribution from Gaussian to uniform. The single-sided lens profile in Fig. 3 is obtained by combining the profiles of a beam-collimating surface and a beam-shaping surface. The collimating profile converts the divergent rays into parallel rays and transforms the spherical wavefront radiated by the antenna into a planar wavefront. Using Fermat's principle, the collimating-surface profile is obtained from the following relation:

$$R(\theta) = \frac{(n-1)L}{n\cos\theta - 1} \tag{1}$$

where $n$, $L$, $R(\theta)$, and $\theta$ denote, respectively, the refractive index of the lens, the distance from the single-sided lens to the antenna phase center, the distance from the antenna phase center to an arbitrary point on the lens surface, and the angle between that point and the lens axis. Because the problem is axisymmetric, the phase profile required to transform the Gaussian beam into a flat-top beam is obtained using the geometrical-optics transformation method [17] as follows [18]:

$$\Phi_{bs}(r) = \frac{2\pi}{\lambda d}\int_0^r \left\{ r_{ft}\sqrt{\frac{1-\exp\left[-2\left(\gamma / r_g\right)^2\right]}{1-\exp\left[-2\left(D_l / 2r_g\right)^2\right]}} - \gamma \right\} d\gamma, 0 < r < D_l / 2 \tag{2}$$

where $D_l$, $r_g$, $r_{ft}$, $r$, $d$, $\lambda$, and $\Phi_{bs}(r)$ denote, respectively, the diameter of the single-sided lens, the radius of the incident Gaussian intensity distribution, the radius of the desired uniform output intensity distribution, the radial distance from the center of the single-sided-lens aperture, the distance from the single-sided lens to the output plane, the operating wavelength, and the beam-shaper phase profile. Using the thin-element method [19], the beam-shaping surface profile is obtained as:

$$h_{bs}(r) = \frac{\lambda \Phi_{bs}(r)}{2\pi(n-1)} \tag{3}$$

where $h_{bs}(r)$ denotes the beam-shaping surface profile. Combining the collimating and beam-shaping surface profiles gives the single-sided lens profile as:

$$h_{pcl}(\theta) = \alpha_{opt} h_{bs}(r) + R(\theta)\cos\theta - L \tag{4}$$

where $\alpha_{opt}$ is an optimization parameter determined by full-wave simulation. All variables associated with the single-sided-lens beam-shaping problem are shown in Fig. 3. As indicated in the figure, $R\sin\theta$ can be substituted for the radial-distance variable $(r)$ in the preceding relation. The wavefront incident on the single-sided lens is spherical, and the rays are simultaneously collimated and reshaped as they pass through the lens. In contrast, in the double-sided lens, the divergent rays from the horn are first collimated by passing through the convex-plano section and are then reshaped by passing through the plano-concave section. The distance from the single-sided lens to the horn phase center, $(L)$, is set to 52 mm so that the radius of the incident Gaussian intensity distribution, $\left(r_g\right)$, is 11.9 mm. This radius is obtained by simulating the horn antenna in CST with the integral-equation solver. The remaining design parameters - the single-sided-lens diameter $(D_l)$, lens-to-output-plane distance $(d)$, lens refractive index $(n)$, radius of the desired uniform output intensity distribution $(r_{ft})$, and operating wavelength $(\lambda)$ - are set to 100 mm, 3 m, 1.45, 400 mm, and 3 mm, respectively. Once the optimization parameter $(\alpha_{opt})$ is determined through full-wave simulations, the single-sided lens surface profile is obtained.

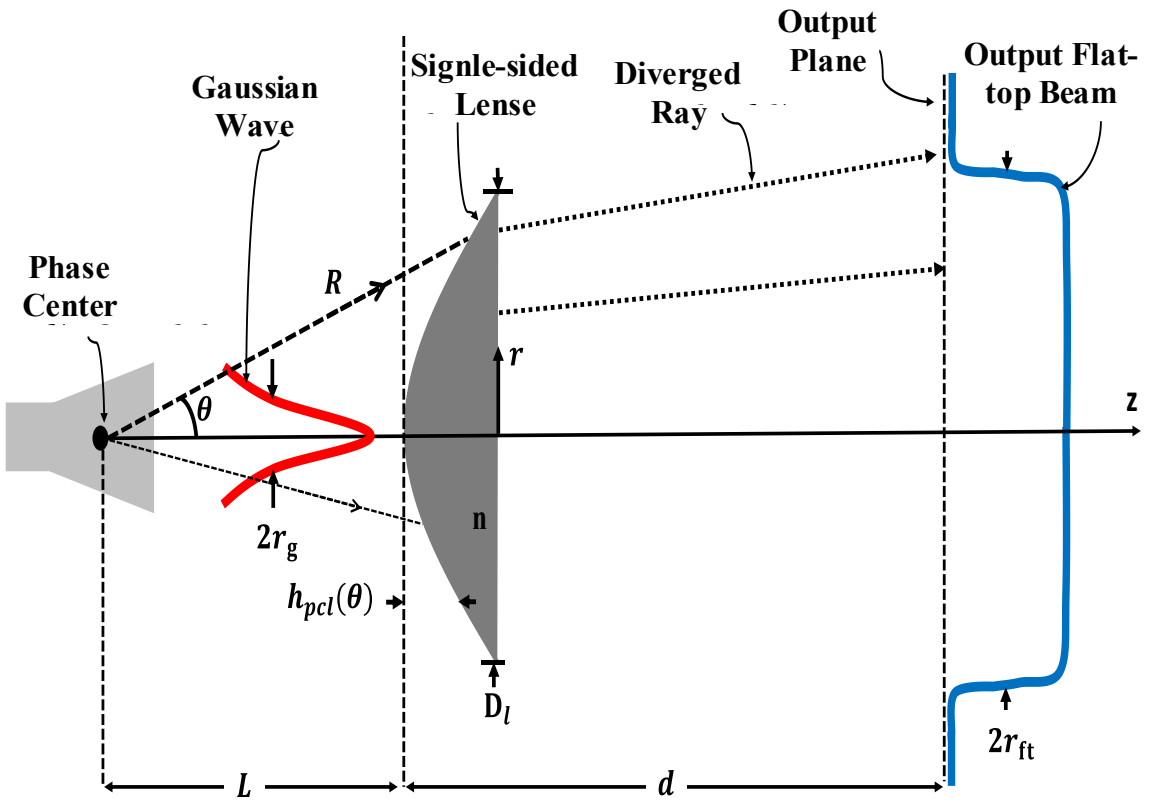

**Fig. 3. Two-dimensional geometry of the single-sided lens and conical horn. The single-sided lens, with diameter $D_l$ and refractive index $n$, is located at a distance $L$ from the antenna phase center and transforms the incident Gaussian intensity distribution with radius $r_g$ into a uniform output intensity distribution with radius $r_{ft}$ at a distance $d$.**

## 4- Full-Wave Simulation of the Single-Sided (Plano-Convex) Lens

To determine the optimization parameter $(\alpha_{opt})$, two performance metrics, ripple and roll-off, are defined. These metrics quantify the quality of the output flat-top intensity distribution. Ripple represents the deviation from the mean intensity within the flat-top beamwidth and is expressed in decibels $(dB)$; roll-off characterizes the steepness of the intensity transition and is calculated from the following relation:

$$\text{roll-off} = 10\log\left(\frac{FWM}{FWHM}\right) \tag{5}$$

where $FWM$ and $FWHM$ denote, respectively, the full width at maximum (FWM) and the full width at half maximum (FWHM). For an ideal flat-top intensity distribution, the ripple and roll-off metrics are $\pm 0dB$ and $0dB$, respectively; for an acceptable flat-top distribution, their values are taken to be below $\pm 1dB$ and above $-1.4dB$, respectively. Because the single-sided lens is electrically large relative to the operating wavelength (3 mm), the integral-equation solver (MLFMM) in CST is used for full-wave simulation of the structure. The horn antenna is excited through a waveguide port, and the optimization parameter is initially set to 1. Full-wave simulation of the horn and the single-sided lens yields an optimized parameter value of 0.79, which satisfies the ripple and roll-off requirements for the desired flat-top intensity distribution. The resulting single-sided lens profile is shown in Fig. 4(a). As seen in the figure, the lens thickness is 42.11 mm. Figure 4(b) compares the normalized output intensity of the single-sided lens with that of the double-sided lens as a function of distance from the center of the output plane. The ripple and roll-off metrics of the single-sided-lens output are also indicated in the figure; their values are $\pm 0.8dB$ and $-1.1dB$, respectively, and satisfy the design requirements. Thus, despite its simplicity, the proposed single-sided-lens design method produces a satisfactory flat-top intensity distribution.

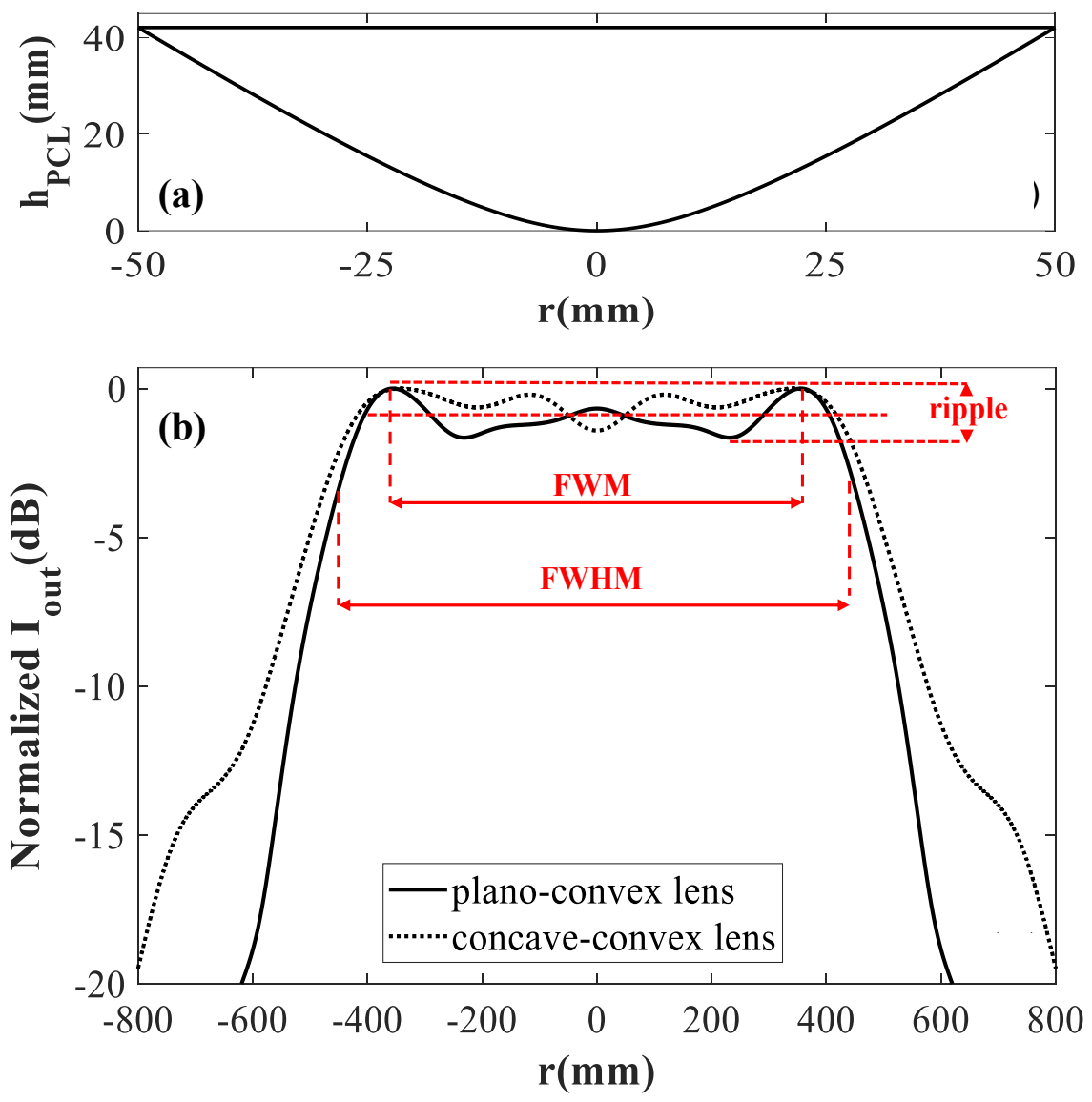

**Fig. 4. (a) Cross section of the single-sided lens surface profile. (b) Normalized output intensity of the single-sided lens (dash-dot line) and the double-sided lens (dashed line) [16] versus distance from the center of the output plane.**

The CST simulation setup used to calculate the normalized electric-field intensity on the output plane in decibels is shown in Fig. 5(a). The output beam uniformly illuminates an approximately circular region with a radius of 0.4 m on a square output plane measuring 1.8 m by 1.8 m. Figure 5(b) shows the three-dimensional normalized output intensity of the single-sided lens.

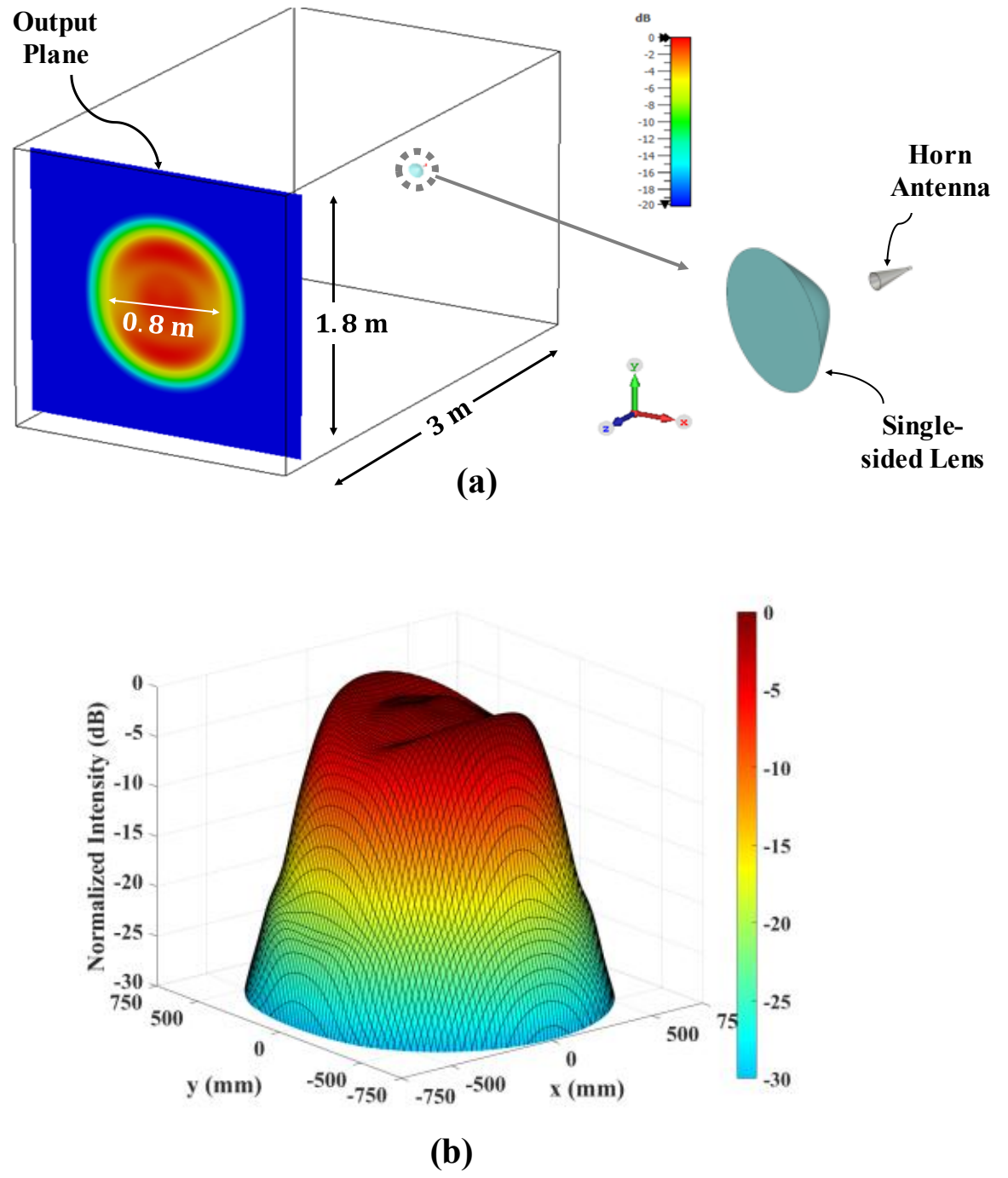

**Fig. 5. (a) Simulation setup of the single-sided lens and millimeter-wave source. (b) Three-dimensional normalized flat-top intensity distribution on the output plane.**

The lens structure designed using (4) is frequency dependent because changing the operating frequency changes both the radius of the incident Gaussian intensity distribution and the wavelength. Consequently, the lens surface profile in (4) changes, which in turn modifies the output flat-top intensity distribution. The output intensities at 90, 100, and 110 GHz for the single-sided and double-sided lenses are shown in Figs. 6(a) and 6(b), respectively. Over the frequency range from 90 to 110 GHz, the output intensity of the single-sided lens maintains satisfactory quality because its beam radius and roll-off vary only slightly; these variations are even smaller than those of the double-sided lens. However, at 90 GHz, the ripple of the single-sided-lens output is larger than that of the double-sided lens and is equal to $\pm 1.08dB$.

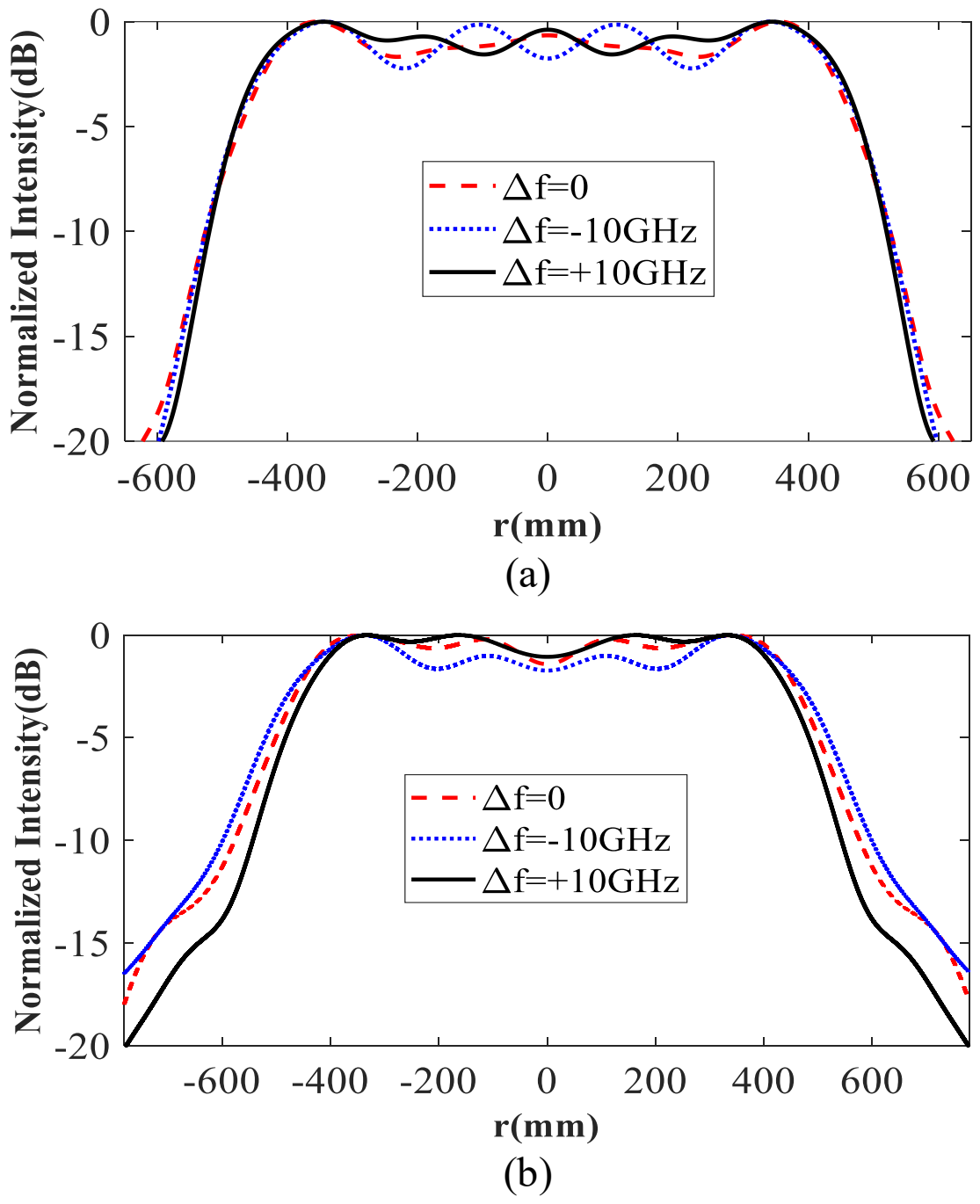

**Fig. 6. Frequency sensitivity of the optimized output intensity for (a) the single-sided lens and (b) the double-sided lens [16].**

## 5- Conclusion

In this paper, a single-sided (plano-convex) lens was designed using a new method to transform the Gaussian beam radiated by a 100 GHz millimeter-wave horn antenna into a uniform intensity distribution at a distance of 3 m. The lens was optimized using full-wave simulations, and the resulting high-quality flat-top distribution validates the proposed approach. The sensitivity of the designed lens to frequency variation was also investigated. As summarized in Table 2, the ripple of the flat-top beam generated by the single-sided lens is relatively sensitive to frequency variation, whereas the roll-off of the output beam is less sensitive. The proposed lens is therefore attractive for modern imaging systems that require uniform illumination of the target plane.

**Table 2. Comparison of the optimized single-sided lens with the optimized double-sided lens [16].**

| | Optimized lens | |
|---|---|---|
| | Double-sided [16] | Single-sided |
| Sensitivity of output-intensity ripple to frequency variation | Lower | Higher |
| Sensitivity of output-beam radius to frequency variation | Higher | Lower |
| Fabrication process | Difficult | Easy |
| Output-intensity roll-off | $-1.3dB$ | $\pm 1.1dB$ |
| Output-intensity ripple | $\pm 0.6dB$ | $\pm 0.8dB$ |